\documentclass[prl,nobalancelastpage,aps,twocolumn,letterpaper,groupedaddress,superscriptaddress, 10.5pt]{revtex4-1}  

\usepackage[hidelinks]{hyperref}

\usepackage{graphicx}  
\usepackage[usenames,dvipsnames]{color}
\usepackage{dcolumn}   
\usepackage{bm}        
\usepackage{amssymb}   
\usepackage{amsmath,textcomp}
\usepackage{xfrac}

\newcommand{\MIT}{\text{LIGO - Massachusetts Institute of Technology, Cambridge, MA}}
\newcommand{\LSU}{\text{Department of Physics, Louisiana State University, Baton Rouge, LA}}
\newcommand{\CMS}{\text{Crystalline Mirror Solutions, Santa Barbara, CA, and Vienna, Austria}}
\usepackage{natbib}

\usepackage[capitalise,nameinlink]{cleveref}
\usepackage[utf8]{inputenc}
\usepackage{float}
\usepackage{easy-todo}
\usepackage{bbold}
\usepackage{mathtools} 
\usepackage{mathrsfs} 
\usepackage{subcaption}
\usepackage{fancyhdr}
\begin{document}

\title{Room temperature optomechanical squeezing}

\author{Nancy Aggarwal}
\email{nancyagg@mit.edu}
\affiliation{\MIT}

\author{Torrey Cullen}
\affiliation{\LSU}

\author{Jonathan Cripe}
\affiliation{\LSU}

\author{Garrett D. Cole}
\affiliation{\CMS}

\author{Robert Lanza}
\affiliation{\MIT}

\author{Adam Libson}
\affiliation{\MIT}

\author{David Follman}
\affiliation{\CMS}

\author{Paula Heu}
\affiliation{\CMS}

\author{Thomas Corbitt}
\email{tcorbitt@phys.lsu.edu}
\affiliation{\LSU}

\author{Nergis Mavalvala}
\email{nergis@mit.edu}
\affiliation{\MIT}

\date{\today}

\maketitle
\thispagestyle{fancy}

\textbf{
The radiation-pressure driven interaction of a coherent light field with a mechanical oscillator induces correlations between the amplitude and phase quadratures of the light. These correlations result in squeezed light -- light with quantum noise lower than shot noise in some quadratures, and higher in others. Due to this lower quantum uncertainty, squeezed light can be used to improve the sensitivity of precision measurements. In particular, squeezed light sources based on nonlinear optical crystals are being used to improve the sensitivity of gravitational wave (GW) detectors. For optomechanical squeezers, thermally driven fluctuations of the mechanical oscillator's position makes it difficult to observe the quantum correlations at room temperature, and at low frequencies. Here we present a measurement of optomechanically (OM) squeezed light,  performed at room-temperature, in a broad band near audio-frequency regions relevant to GW detectors. We observe sub-poissonian quantum noise in a frequency band of 30 kHz to 70 kHz with a maximum reduction of 0.7 $\pm$ 0.1 dB below shot noise at 45 kHz. We present two independent methods of measuring this squeezing, one of which does not rely on calibration of shot noise.
}

Measurements whose sensitivity is limited by quantum noise can be improved by modifying the distribution of quantum noise. For example, the shot noise limit of optical measurements using coherent states of light can be improved by using squeezed states \cite{Caves1981,Caves_1980,G&K}. Squeezing methods employ light with uncertainty below shot noise in the signal quadrature at the expense of increased noise in orthogonal quadratures. As a result of this redistribution of uncertainty, squeezed states can enhance the precision of measurements otherwise limited by quantum noise. The preeminent example is interferometric GW detectors where squeezed state injection lowers the noise floor below shot noise \cite{KLMTV,Caves1981,Caves_1980,LIGO_sqz,GEO_sqz}. 

Squeezed states of light suitable for GW detectors have been successfully generated using nonlinear optical materials \cite{Wu1986,LIGO_sqz,GEO_sqz,Schnabel2010,Schnabel2017}. The OM interaction is similarly an effective nonlinearity \cite{RevModPhys.86.1391} for the light field, which can squeeze its quantum fluctuations \cite{PDSQ1,PDSQ2,KLMTV,Harms,Corbitt_OM_Squeezing}. Optomechanical squeezing has some advantages over squeezed state generation using nonlinear optical media. OM squeezing can be generated independent of the optical wavelength, with a tunable frequency dependence of the squeezing quadrature via the optical spring \cite{Corbitt_OM_Squeezing,Aggarwal_OptOpt}, and in the long term OM squeezers have great potential to be miniaturized. 

Previously, OM squeezing has been observed \cite{DSK_Sq,Painter_Sq,Regal_StrongSq,Sudhir_Sq} in systems operated close to the mechanical resonance (within an octave). While these experiments laid important foundations for OM squeezed light, some important challenges for practical OM squeezed light sources remained. For GW detection, for example, the squeezed light source needs to be broadband over three decades in the audio-frequency band, compact, and operating stably 24/7 at room temperature. Here, we present a measurement of squeezing produced by an OM system comprising a Fabry-Perot interferometer with a micro-scale mirror as a mechanical oscillator at room temperature, where for the first time OM squeezing has been observed in a room temperature system, at frequencies as low as tens of kilohertz and extending more than a decade away from the mechanical resonance. This observation of broadband OM squeezing at room temperature presents a new avenue for building quantum OM resources at room temperature that are independent of laser wavelength. 

Overcoming thermal noise \cite{Saulson} has been a fundamental challenge to observing optomechanically generated squeezing beyond cryogenic temperatures. Reducing the quantum noise below shot noise in such a system is only possible if the motion of the oscillator has a significant contribution from quantum radiation pressure noise (QRPN), and is not overwhelmed by thermal fluctuations \cite{Aggarwal_OptOpt,Aggarwal_MechOpt}.
Our mechanical oscillators are designed to have extremely low broadband thermal noise \cite{Aggarwal_MechOpt,cole08, cole12,  cole14, Singh_PRL} and have been used to observe QRPN \cite{Cripe_QRPN}. The thermal noise of these oscillators is sufficiently low to not overwhelm the effect of QRPN. Even so, thermal noise does limit the amount of measurable squeezing generated. In addition to the limitation set by thermal noise, our locking and detection scheme introduces losses that degrade some of the quantum correlations created by the OM coupling. Thermal noise, lossy detection, and cavity-feedback noise together limit the amount of squeezing and the frequency band in which it is observed.

A precise calibration of shot-noise has been the basis for all prior demonstrations of optomechanical squeezing. We demonstrate a technique based on photocurrent correlations that obviates the need for such a calibration. This technique may be useful on its own for future studies of squeezing in general.

\begin{figure}
	\includegraphics[width=\columnwidth]{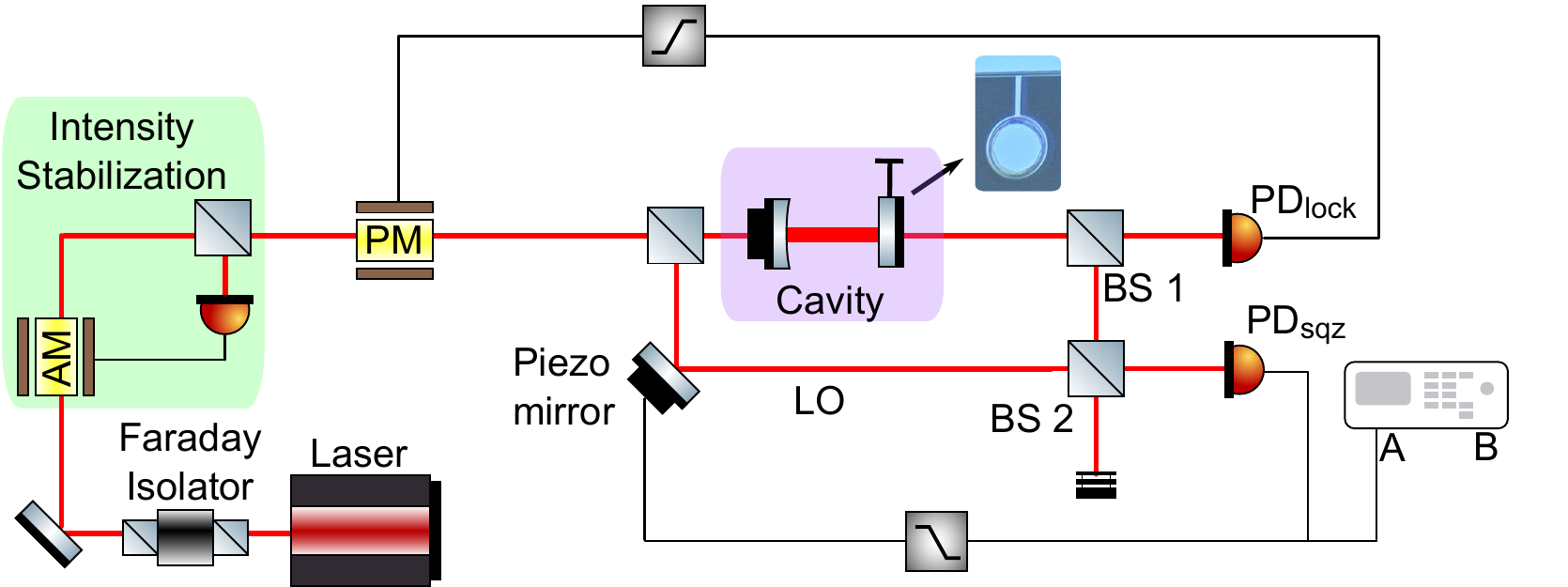}
	\caption{An overview of the main subsystems in the experiment. First, the classical intensity noise of an 1064 nm Nd:YAG laser is suppressed by an intensity stabilization servo using an amplitude modulator (AM) as the actuator. The light is then sent to an optomechanical cavity -- the input mirror is a mechanically rigid macro-mirror, and the output mirror is a low-noise micro-scale mirror supported by a single-crystal micro-cantilever. The light inside this cavity gets squeezed due to the radiation pressure interaction between the circulating light and the movable micro-mirror. The cavity is locked by picking off 15\% of the transmitted power through $\mathrm{BS}_1$ on $\mathrm{PD}_\mathrm{lock}$, and feeding back that signal to a phase modulator (PM). The remaining 85\% of the light is interfered with a local oscillator on $\mathrm{BS}_2$ which reflects 96.5\% and transmits 3.5\% of the light. The phase between the LO and signal is locked by feeding back the DC part of the fringe detected on $\mathrm{PD}_\mathrm{sqz}$ to a piezo mirror in the LO path. The signal from $\mathrm{PD}_\mathrm{sqz}$ is also sent into the spectrum analyser for measurement.}
	\label{fig:Setup}
\end{figure}

Our experimental setup consists of two main subsystems -- the optomechanical cavity and the detection system, as shown in \cref{fig:Setup}. The optomechanical system is a Fabry-Perot cavity pumped with a 1064 nm Nd:YAG NPRO laser. One of the two mirrors of this cavity is supported by a low-noise single-crystal micro-cantilever (similar to that employed in \cite{Cripe_QRPN}), with a mass of 50 ng, a fundamental frequency of 876 Hz, and a mechanical quality factor of 16,000. The other mirror is a 0.5 inch diameter mirror with radius of curvature 1 cm. The cavity is just under 1 cm long, has a finesse of around 11,500, and a HWHM linewidth ($\gamma$) of 650 kHz. 

We lock the cavity slightly blue detuned, at around 0.33$\gamma$ away from resonance by using the strong optical spring (145 kHz) created by the detuned operation \cite{Corbitt_OM_Squeezing}. The optical spring by itself is unstable, so an electronic feedback in frequencies near the optical spring is used to stabilize the system using the transmitted light for the error signal. We use radiation pressure as the actuator for locking, as detailed in detail in Ref. \cite{Cripe_RPL}, with one difference: in this experiment, we use a phase modulator as our actuator instead of an amplitude modulator. We can treat the instability of the optical spring in the same way, except for a slightly modified plant transfer function \cite{Aggarwal_ClosedLoop}. The open loop gain of the cavity locking loop is below one at all frequencies less than 140 kHz. Since we must obtain a signal to stabilize the optical spring while leaving the squeezed light available to be independently measured, we split the light exiting the cavity at a beam-splitter ($\mathrm{BS}_{1}$), using 15\% of the total light to obtain the feedback error signal. This method introduces some common phase noise between the local oscillator and cavity field, which is included in our noise budget.

Traditionally, balanced homodyne detection is used to characterize squeezing, since it cancels classical intensity noise of the local oscillator and does not introduce loss. In our setup, however, we use a different method to measure the squeezing. This is because the classical intensity noise is sufficiently small to not require cancellation, and the level of squeezing we expect is low, making it insensitive to a small loss. The beam transmitted from the cavity (signal) is combined with a local oscillator (LO) beam on a 96.5\%-3.5\% beam splitter ($\mathrm{BS}_2$), as shown in \cref{fig:Setup}. We then measure the port that has 96.5\% signal and 3.5\% LO on a photodetector ($\mathrm{PD_{sqz}}$). The output of $\mathrm{PD_{sqz}}$ is low-pass filtered, amplified, and then fed back to a piezoelectric crystal driving the length of the LO path. This loop suppresses relative path length fluctuations between the signal and the LO, but only at frequencies well below the measurement band. The loop has a unity gain frequency of less than 1 kHz, and has an open loop gain of less than -40 dB at the measurement frequencies, as shown in \cref{fig:HomodyneLoop} in the Supplemental Information (SI). This eliminates the need to correct the squeezing spectrum for the response of the feedback loop.  Additionally, there is no cross over between the homodyne loop and the cavity loop because their frequency regions of actuation are disjoint. Note that since $\mathrm{PD_{sqz}}$ is an out-of-loop detector for both the cavity-locking as well as the homodyne-locking loop, a sub-shotnoise measurement on it is an indication of squeezing \cite{Wiseman99}. The lock maintains $\mathrm{PD_{sqz}}$ at a constant DC voltage level, which we use to calibrate the shot noise level. The measurement quadrature is determined by the relative path length between the signal and LO. In the laboratory, the measurement quadrature can be tuned by changing either the lockpoint level, or the LO power, or both, see \cref{fig:phasor,eq:MeasurementQuadrature} in SI.

\begin{figure}
	\includegraphics[width=\columnwidth]{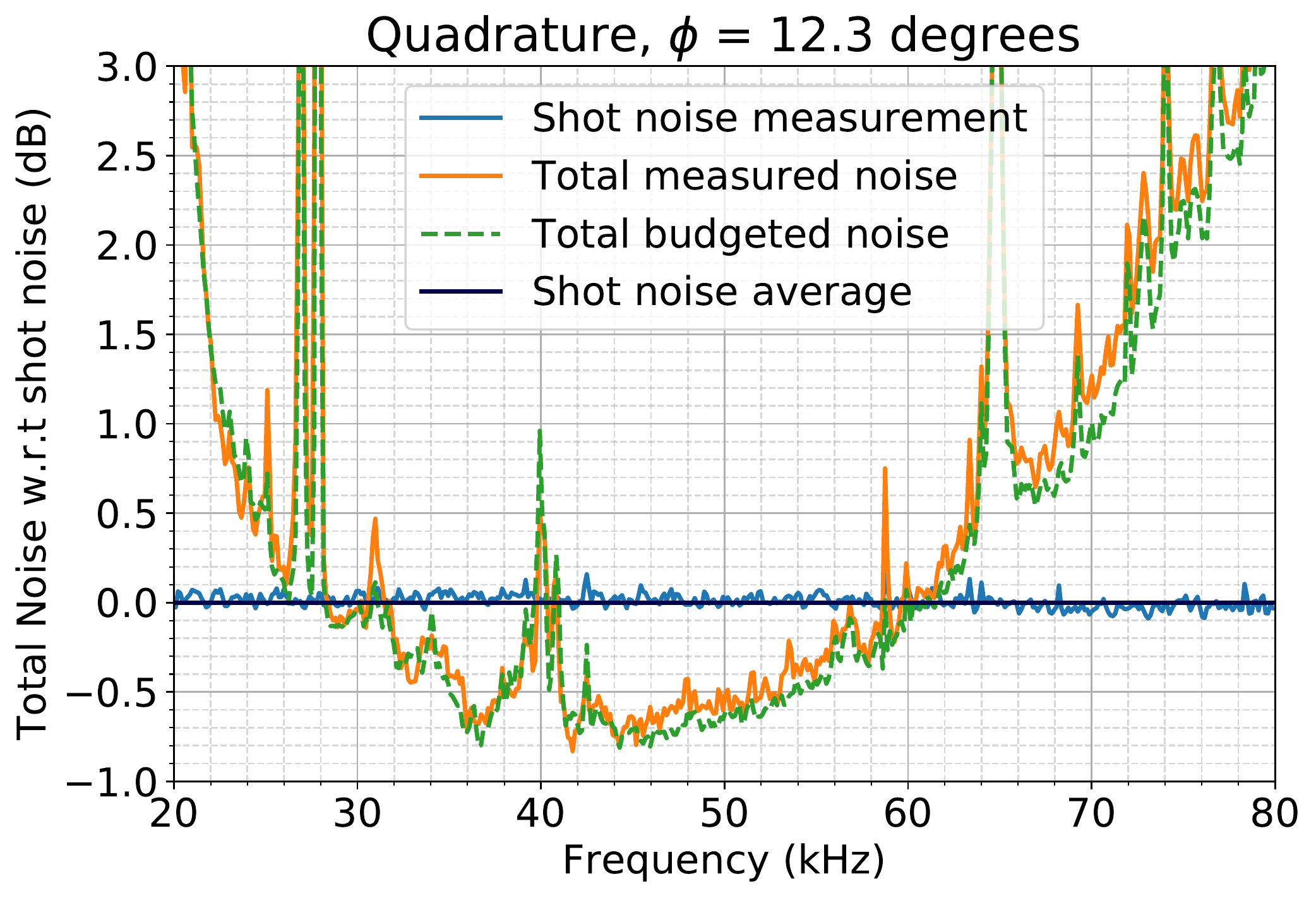}
	\caption{Measured spectrum and modeled noise budget at 12.3 degrees quadrature.  All quadrature angles are referenced so that 0 degrees corresponds to the amplitude quadrature of the cavity transmission. This figure shows the measured  spectrum relative to shot noise. We show the shot noise measurement in blue, which is used to obtain an average shot noise level. All the data in the paper is scaled to this average shot noise level. The  spectrum for total measured noise at 12.3\textdegree \hspace{0.01cm} is shown in orange, showing squeezing from 30 kHz to 60 kHz, with maximum squeezing of 0.7 $\pm$ 0.1 dB (corresponding to a 15 $\pm$ 2\% reduction in PSD) near 45 kHz. We also show  the total budgeted noise  in dashed green which is a quadrature-sum of  quantum noise, thermal noise, classical laser noise, cavity-feedback noise, and differential phase noise.} 
	\label{fig:Spectrum}
\end{figure}

In order to compare the measured noise to shot noise, we measure the shot noise level by turning off the homodyne lock, blocking the signal port, and tuning the LO power to get the same voltage on $\mathrm{PD_{sqz}}$ as our lockpoint. This allows us to measure a spectrum of $\mathrm{PD_{sqz}}$ that contains shot noise of the light, classical intensity noise, and the dark noise of $\mathrm{PD_{sqz}}$. We then average this spectrum over our measurement band to obtain the reference level (0 dB). Classical relative intensity noise (RIN) is suppressed by an intensity stabilization servo (ISS) to about $8 \times 10^{-9} / \sqrt{\mathrm{Hz}}$, and contributes less than $-40$ dB of the noise on $\mathrm{PD_{sqz}}$. The RIN level is independently measured by performing a correlation measurement between $\mathrm{PD_{sqz}}$ and another pick-off between the ISS and the PM. Dark noise accounts for about -12 dB of the shot noise level, and is not subtracted. 

The result of the homodyne measurement of the signal is shown in Fig. \ref{fig:Spectrum}. For a quadrature angle of 12.3\textdegree \hspace{0.01cm} from the amplitude quadrature, we observe up to 0.7 $\pm$ 0.1 dB of squeezing (equivalent to a 15 $\pm$ 2 \% reduction in the PSD), from 30 kHz to 60 kHz.  

\begin{figure}
	\includegraphics[width=\columnwidth]{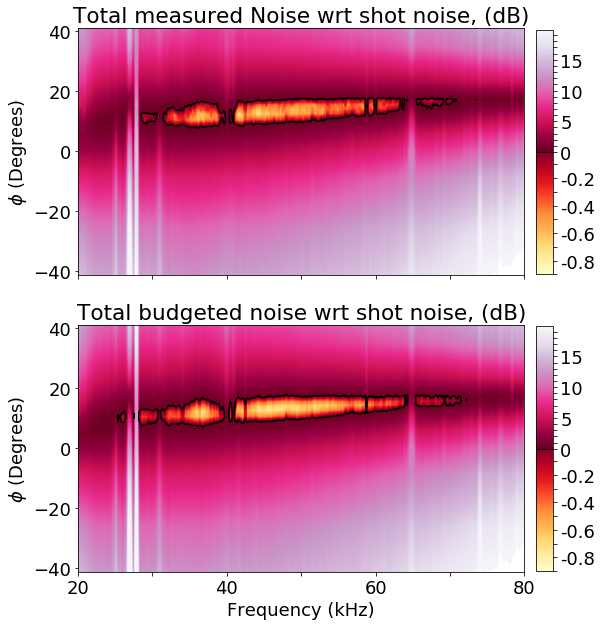}
	\caption{\textbf{Top:} Measured noises on $\mathrm{PD}_\mathrm{sqz}$ at fourteen different quadratures, distributed more densely near the squeezing quadrature, and sparsely elsewhere. The black contour line corresponds to shot noise. The regions inside it are squeezed (shown by the lower colorbar), and the region outside it are antisqueezed (shown by the upper colorbar). We observe squeezing from 10\textdegree \hspace{0.01cm} to 17\textdegree \hspace{0.01cm} and from 30 kHz to 70 kHz. We can also see one of the mechanical modes of the cantilever at 27 kHz. \\
	\textbf{Bottom:} Budgeted noise relative to shot noise. The color scheme is same as the top panel. As is characteristic of OM squeezing below the optical spring frequency \cite{Aggarwal_OptOpt}, the higher quadrature shot noise crossing for all frequencies occurs at the same quadrature. The upper part of the shot noise contour is nearly perfectly horizontal.}
	\label{fig:Contour}
\end{figure}

The distribution of squeezing is studied in detail by measurements of other quadratures of the homodyne signal.  In order to do this without changing the locking loop or shot noise, we keep the homodyne locking offset the same, and vary the LO power. This allows us to change the measurement quadrature in a shot noise invariant way. In the top panel of \cref{fig:Contour}, we show this measurement as a function of sideband frequency and quadrature.

To understand the observed squeezing, a detailed noise budget of the system is developed. The total budgeted noise in the squeezing quadrature is shown in \cref{fig:Spectrum}, and in a quadrature dependent way in the bottom panel of \cref{fig:Contour}. This noise budget includes a model \cite{Corbitt_mathematical} that predicts the contribution of quantum noise and previously measured thermal noise \cite{Cripe_QRPN} for the measured cavity and homodyne parameters; measured cavity-feedback injected noise and differential phase noise between the LO and the cavity. Finally, the extra loss in the detection path is obtained by comparing measurement and noise budget at all frequencies and quadratures. Further details on the noise budget can be found in the SI. As we see, the overall behavior of the system is similar in the measurement as well as noise budget, most importantly the squeezing quadrature. 
 
\begin{figure}
	\begin{subfigure}{\linewidth}
		\centering
		\includegraphics[width=\linewidth]{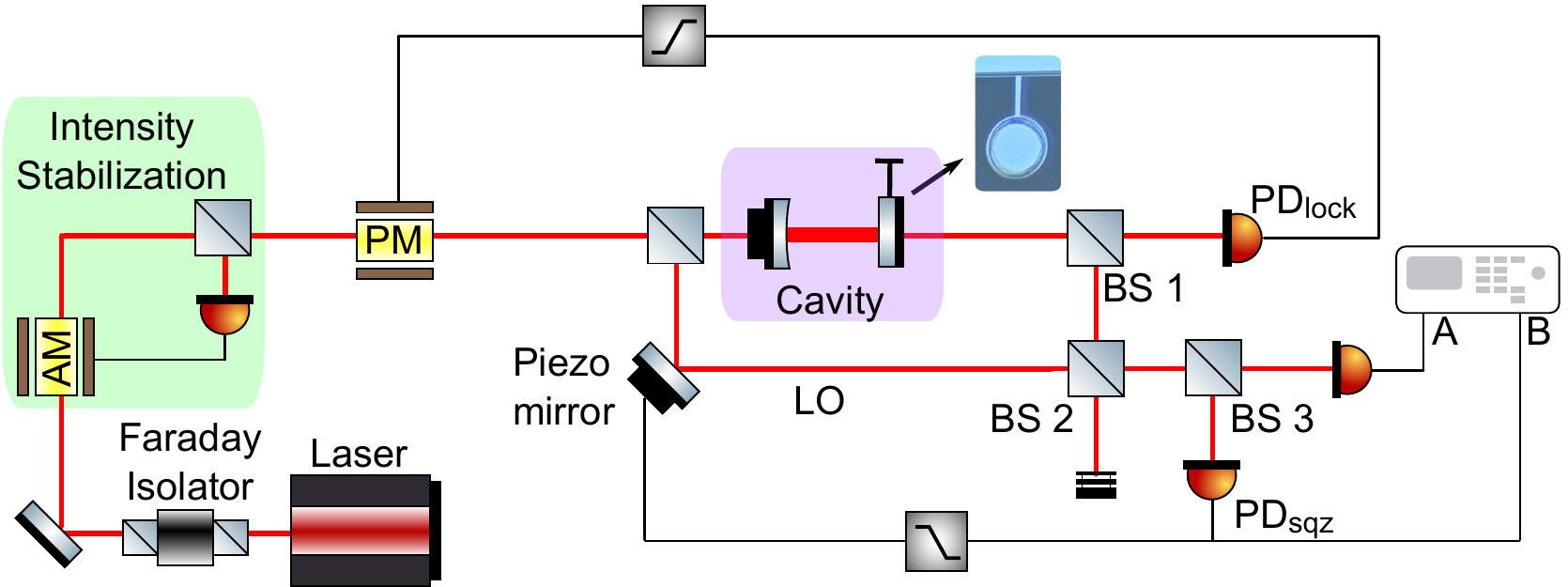}
		\caption{}
		\label{fig:CorrelationSetup}
	\end{subfigure}
	\begin{subfigure}{\linewidth}
		\centering
		\includegraphics[width=\columnwidth]{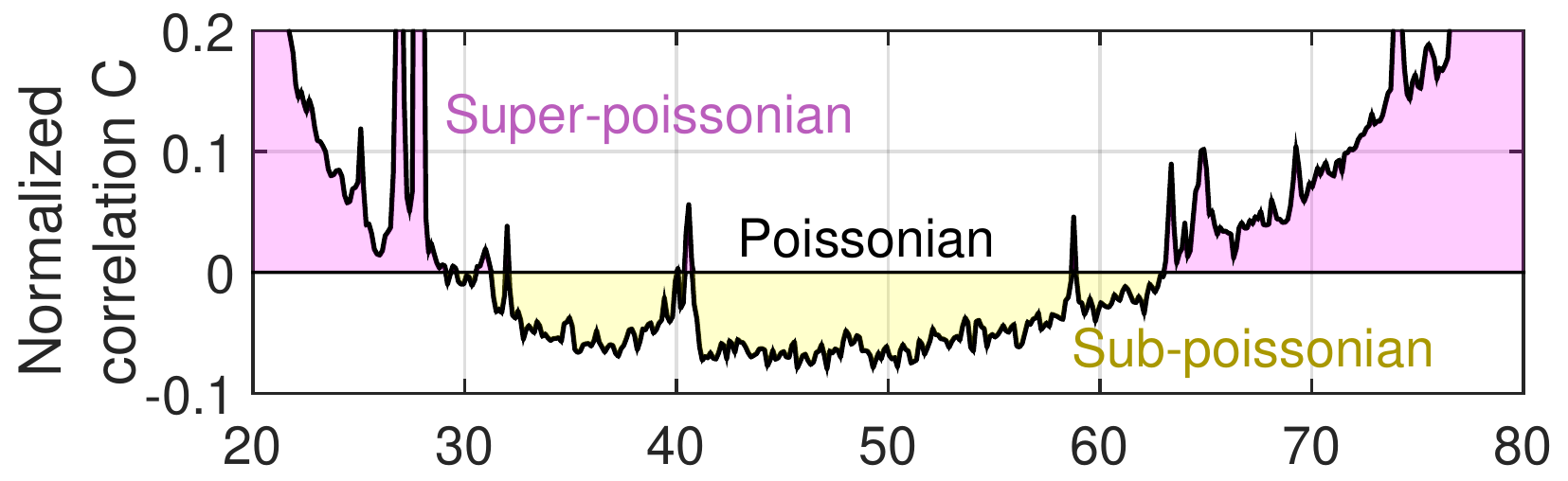}
		\caption{}
		\label{fig:Correlation}
	\end{subfigure}
	\begin{subfigure}{\linewidth}
		\centering
		\includegraphics[width=\linewidth]{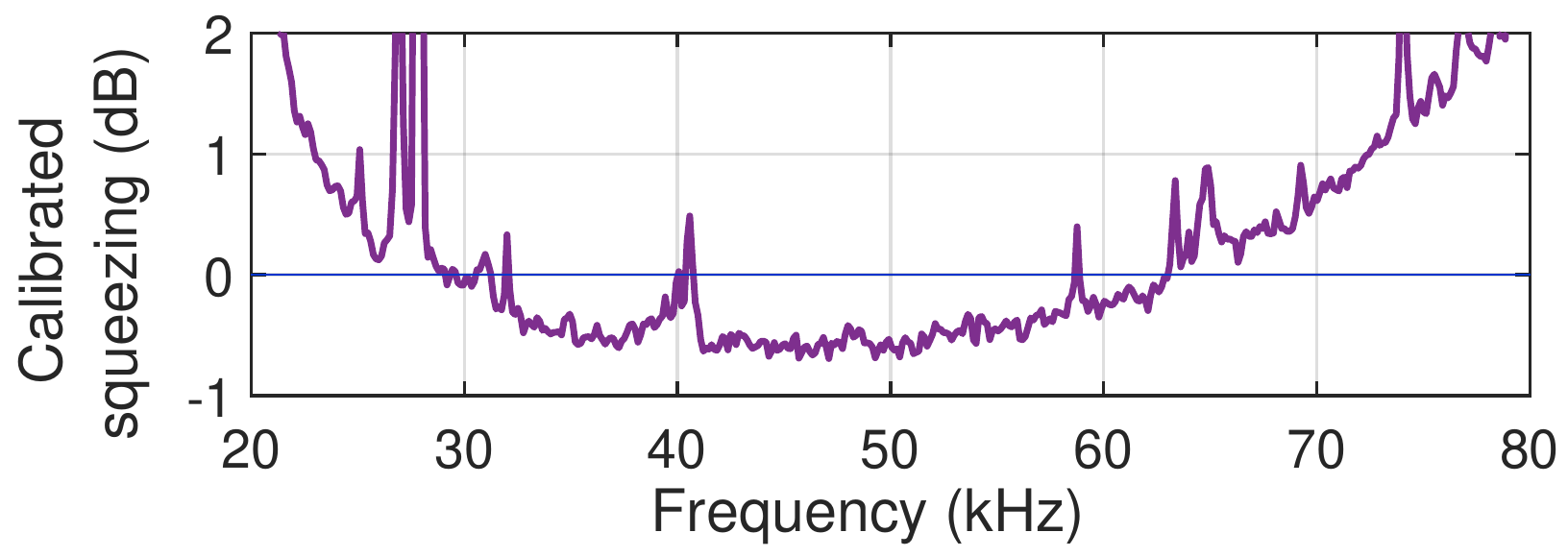}
		\caption{}
		\label{fig:Correlation_squeezing}
	\end{subfigure}
	\caption{\textbf{(\subref{fig:CorrelationSetup})}  Setup for correlation measurement: We set the LO such that the field after $\mathrm{BS_2}$ is amplitude squeezed and pass it through a 50-50 ($\mathrm{BS_3}$). We then perform a cross-spectrum measurement of the two outputs and normalize it to the individual spectra. This quantity can only be negative if the input beam is squeezed in the amplitude quadrature (see \cref{eq:Sab} in SI). \textbf{(\subref{fig:Correlation})} Measurement of negative correlations. The existence of these negative correlations provides a verification of squeezing, and allows for a shot-noise independent way of verifying the existence of amplitude squeezing. \textbf{(\subref{fig:Correlation_squeezing})} Squeezing spectrum obtained by using the negative correlations shown in (\subref{fig:Correlation}) (see \cref{eq:C_to_R} in SI). This spectrum is inherently calibrated to shot noise.}
	\label{fig:Correlation_Setup_And_Measurement}
\end{figure}
	
For additional evidence of squeezing, we have also performed a correlation measurement on the squeezed light. Extending the approaches in Refs. \cite{Bartley_Sub-Binomial,Kimble_antibunching,Lee_Note,Krivitsky_Correlation,PhysRevLett.59.2555}, we demonstrate that these correlations are a way to characterize a squeezed light source without measuring shot noise. The light exiting the cavity, after combination with the LO, is split equally between two photodetectors, as shown in \cref{fig:CorrelationSetup}. As described in the SI, if the light is limited by classical noise, positive correlations should be observed in the two photocurrents. Shot noise limited light should produce zero correlations, and intensity squeezed light should produce negative correlations. We measure the cross power spectrum between the two photodetectors and confirm that negative correlations are observed, as shown in \cref{fig:Correlation}. The cross-spectrum is negative from 33 kHz to 62 kHz, and positive elsewhere, which agrees with the measured spectrum in \cref{fig:Spectrum}. For explicit comparison, we convert this correlation to the squeezing factor (using \cref{eq:C_to_R} in SI). This squeezing factor is shown in \cref{fig:Correlation_squeezing}. This provides unconditional evidence that the light is squeezed at these frequencies and at this quadrature. 

To summarize, we report the first observation of room temperature, broadband optomechanical squeezing at frequencies as low as 30 kHz. This measurement not only paves the way for building miniature, wavelength-agnostic devices to improve the performance of quantum measurements like GW measurements, but also opens up possibilities of exploring broadband quantum correlations at room temperature.

\begin{acknowledgments}
	This work was supported by the National Science Foundation grants PHY-1707840, PHY-1404245, PHY-1806634 and PHY-1150531. We are particularly grateful to Vivishek Sudhir, Lee McCuller and Min Jet Yap for valuable discussions and for their detailed comments on this manuscript. The micro-resonator manufacturing was carried out at the UCSB Nanofabrication Facility. We also thank MathWorks, Inc for their computing support. 
\end{acknowledgments}

\bibliographystyle{unsrt}
\bibliography{biblio_Squeezing}
\clearpage


\section{Supplementary Material}

\subsection{Application to precision measurements}

In addition to being all-wavelength and compact, an OM squeezer innately generates a bright squeezed field. This means an OM squeezed field comes with its own internal phase reference, hence eliminating the need for an extra coherent phase locking field (CLF) \cite{Oelker_phase_noise,LIGO_sqz}. For instance, if the state after $\mathrm{BS_2}$ was sent to a precision measurement setup, it would include the carrier field which provides a self-referenced signal to allow locking to the correct squeezing quadrature. In contrast, vacuum squeezed states generated by a nonlinear crystal have to be accompanied by an extra, usually frequency shifted optical beam that keeps track of the squeezing quadrature \cite{Oelker_phase_noise,LIGO_sqz}.

\begin{figure}[h!]
	\includegraphics[width=\linewidth]{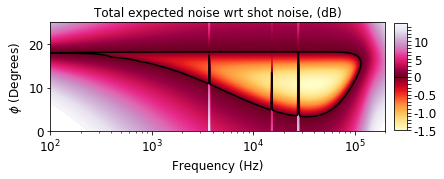}
	\caption{Expected squeezing with lower loss and in the absence of technical noises. The differential phase noise masks the squeezing at low frequencies, whereas the noise injected by the cavity feedback electronics degrades the high frequency side of the correlations. Once these technical noises have been suppressed, and the optical losses have been lowered, we would expect to see about 1.5 dB of squeezing from this system. This limit comes from a combination of escape efficiency and thermal noise \cite{Aggarwal_OptOpt}.}
	\label{fig:ExpectedSqueezing}
\end{figure}

While the current amount of squeezing in our setup is limited by extraneous technical noises, in the absence of those noises, the squeezing would be limited by intracavity losses. This expected squeezing in the absence of extraneous noise is shown in \cref{fig:ExpectedSqueezing}. After elimination of differential phase noise between the LO and cavity, the squeezing would extend to much lower frequencies. The differential phase noise injected into the system can be reduced by picking-off and recombining the LO inside the vacuum where the OM cavity is situated. The feedback noise can be subtracted by monitoring the actuator signal. A more optimized feedback scheme could also be implemented that injects lower noise, e.g. a feedback tuned for the measured quadrature's transfer function with respect to an appropriate input quadrature that would couple minimum noise to the squeezing signal \cite{Aggarwal_ClosedLoop}. 

\subsection{Noise budget}

The measurement is also compared to a noise budget, shown in the bottom panel of \cref{fig:Contour}. The total budgeted noise shown is the quadrature sum of individual contributions from measured thermal noise, quantum noise, classical laser noises, cavity-feedback noise, and differential phase noise between the signal and LO.	It uses experimentally measured cavity parameters, thermal noise, $\mathrm{BS_1}$ and $\mathrm{BS}_2$ reflectivities and homodyne visibility, listed in \cref{table:ExptParams}. The quadratures for which squeezing is obtained depend on the various OM parameters of the cavity, such as the detuning, circulating power, losses, as well as the thermal noise.  We measure the cavity detuning, intracavity power and losses by measuring transfer functions from amplitude modulations on input to transmitted light. The thermal noise is measured by a cross-spectrum measurement in the amplitude quadrature without the local oscillator \cite{Cripe_bae}. We have also separately calibrated all the beam splitters, the mirror reflectivities, and the homodyne visibility. These measured quantities are then used to predict the squeezing using a numerical model based on Ref. \cite{Corbitt_mathematical}. In this model, we also include the effect of the unbalanced homodyne with an imperfect visibility and common-mode laser noises. 

We then characterize the impact of technical noises by measuring their contribution. First, we measure the contribution of noise injected by the cavity locking system. The dominant source of this noise is shot noise at $\mathrm{PD}_{\rm lock}$ due to 15\% transmission of $\mathrm{BS_1}$. In order to measure this feedback noise, we measure the coherence between $\mathrm{PD}_{\rm sqz}$ and the amplifier output that is fed to the phase modulator at input. This coherence when multiplied with the spectrum of $\mathrm{PD}_{\rm sqz}$ gives us the contribution of feedback noise. We do this at all measurement quadratures independently. We find that the impact of feedback noise is minimized at 17 degrees, akin to other intracavity displacement noises like thermal noise. 

In principle, this cavity-feedback noise could be subtracted from the final result, as it is a measured quantity, but we choose not to do so for the sake of simplicity. Instead, we chose to pick-off the LO beam just after the cavity-feedback phase modulator, so that there is common mode rejection of this locking loop phase noise at the homodyning stage at $\mathrm{BS}_2$. The common mode rejection by the homodyne detection also allows us to cancel frequency noise originating from the NPRO laser, without requiring a frequency stabilization servo. Any scheme to measure squeezing not purely in the amplitude quadrature requires mixing the signal beam with an LO that is phase coherent with it, and so one always has the ability to reject common mode noise in this fashion. Also note that there is no risk of generating apparent squeezing after $\mathrm{BS}_2$ by deriving the LO from the cavity locking field (e.g. from feedback-squashing of the in-loop field), because the LO and signal fields are both out-of-loop \cite{Wiseman99}.

Additionally, displacement fluctuations that are relative between the LO and the signal path cause an effective phase noise in the measurement. We refer to this as the differential phase noise, and we measure it by analyzing the measured noise at 17\textdegree. At this quadrature, all displacement noises including the feedback noise are canceled, and the quantum noise contribution is at the shot noise level. So we attribute all noise above shot noise at 17 degrees to this relative phase noise. We calculate the contribution of phase noise in all other quadratures by assuming that it is maximum at 90\textdegree quadrature and scaling it sinusoidally. 

\begin{figure}
	\begin{subfigure}{\linewidth}
		\centering
		\includegraphics[width=\linewidth]{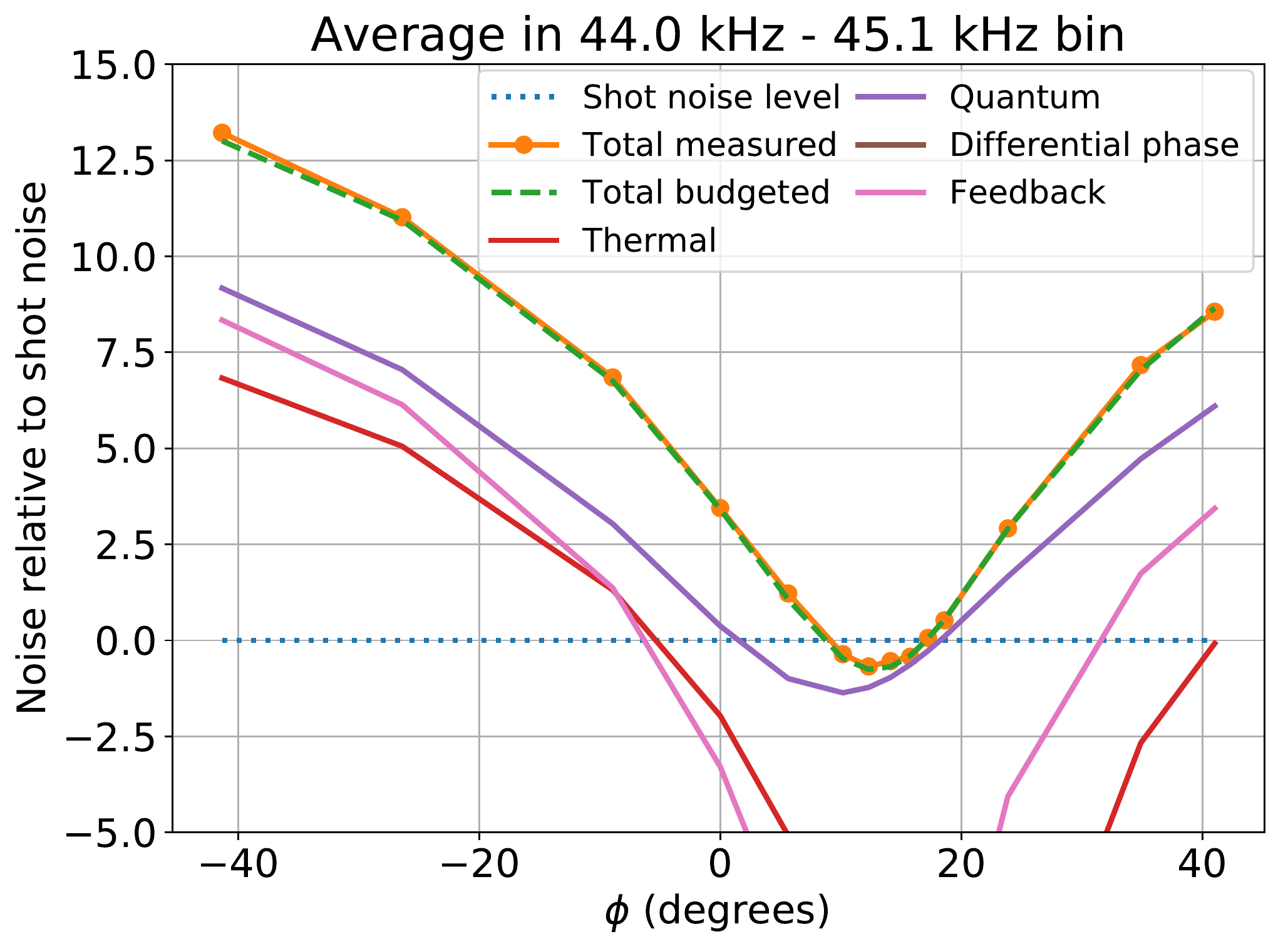}
		\caption{}
		\label{fig:McD}
	\end{subfigure}
	\begin{subfigure}{\linewidth}
		\centering
		\includegraphics[width=\linewidth]{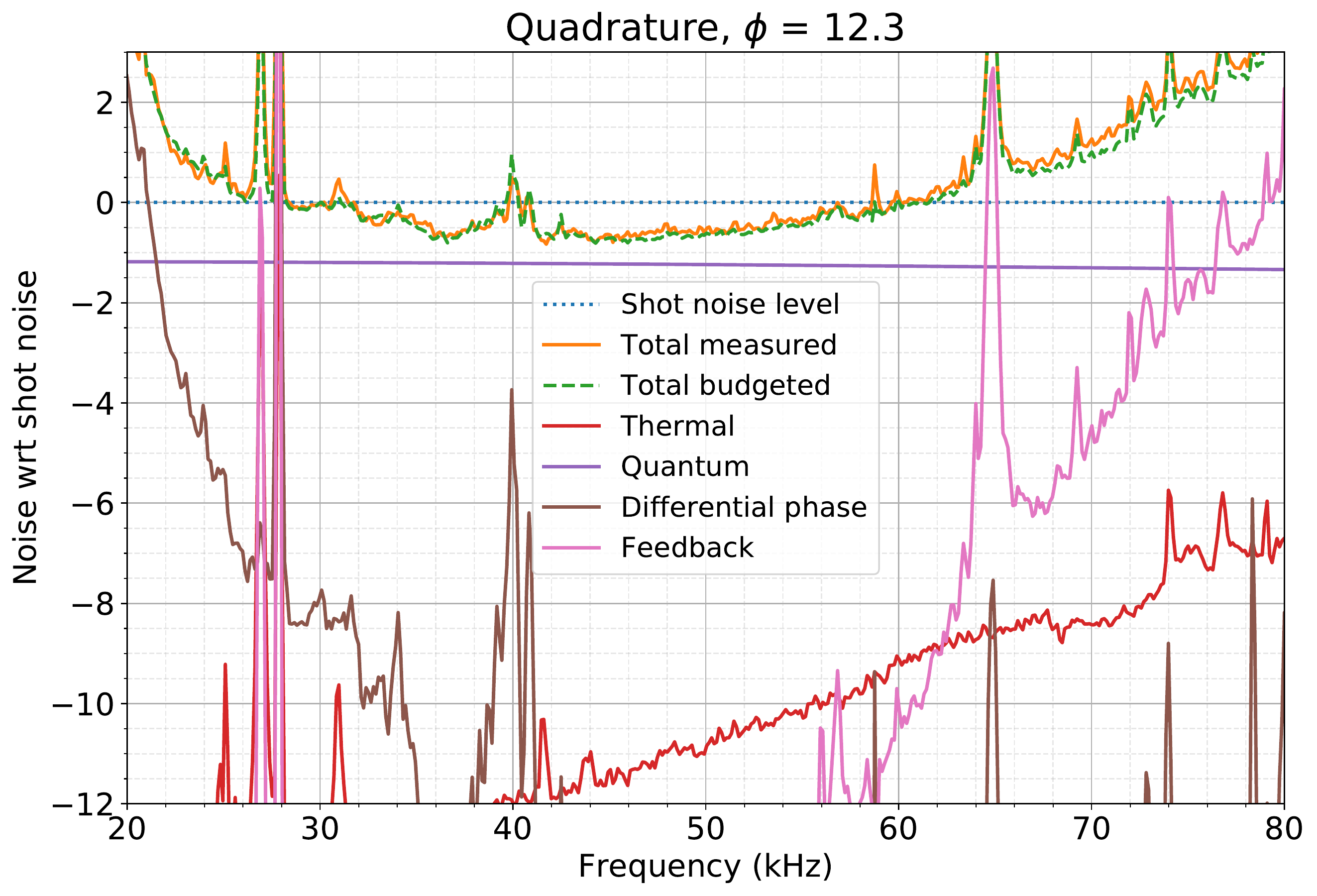}
		\caption{}
		\label{fig:Noise_Budget_12.3deg}
	\end{subfigure}
	\caption{Noise budget: \textbf{(\subref{fig:McD})} Measurement and noise budget as a function of quadrature, averaged over a 1 kHz bin. \textbf{(\subref{fig:Noise_Budget_12.3deg})} Measurement and full noise budget at the squeezing quadrature. The measured noise at 12.3\textdegree is shown in orange. Also shown are the contributions from quantum noise (with excess loss) in purple, thermal noise in red, differential phase noise in brown, and cavity-feedback noise in pink. The quadrature sum of all these contributions is shown in dashed green. All noises are relative to shot noise and are shown in dBs.}
	\label{fig:Ind_Noise_Budgets}
\end{figure}

Finally, we are left with excess loss in the detection path. We fit this loss by adding a frequency and quadrature independent loss to the noise budget. We find an excess loss of 22 $\pm$ 1\%, which agrees with the measured loss of 21 $\pm$ 8\%. Note that optical loss is the only effect where a single scalar would be sufficient to explain the measurement over all quadratures and frequencies. All of the above contributions to the noise budget are shown in \cref{fig:Ind_Noise_Budgets}: as a function of measurement quadrature in \cref{fig:McD}, and as a function of frequency in the squeezing quadrature in \cref{fig:Noise_Budget_12.3deg}.

\begin{table}
	\centering
	\begin{center}
	\begin{tabular}{||l|l||}
		\hline
		$\mathrm{BS}_1$ reflectivity & 85 \% \\
		$\mathrm{BS}_2$ reflectivity & 96.5 \% \\ 
		Input coupler transmission & 50 ppm\\ 
		Cantilever mirror transmission & 250 ppm \\
		Cavity losses & 250 $\pm$ 20 ppm\\ 
		Cavity linewidth HWHM ($\gamma$) & 650 kHz \\ 
		Cavity detuning & 0.33$\gamma$\\ 
		Homodyne visibility & 0.93 \\
		Intracavity power & 260 $\pm$ 30  mW\\
		Signal power $\left(\left|t \mathrm{\vec{E}_S}\right|^2\right)$ & 58 $\pm$ 4  \textmu W \\
		LO power $\left(\left|r \mathrm{\vec{E}_{LO}}\right|^2\right)$  & 0 - 30 $\pm$ 3 \textmu W   \\
		Detected power $\left(\left|\mathrm{E_{sqz}}\right|^2\right)$ & 49 $\pm$ 3 \textmu W\\
		Detection inefficiency and extra losses & 21 $\pm$ 8 \% \ \\
		\hline
	\end{tabular}
	\end{center}
	\caption {Experimental parameters determined from measurements in the lab}
	\label{table:ExptParams}
\end{table}

\begin{figure}[h!]
	\centering
	\includegraphics[width=\linewidth]{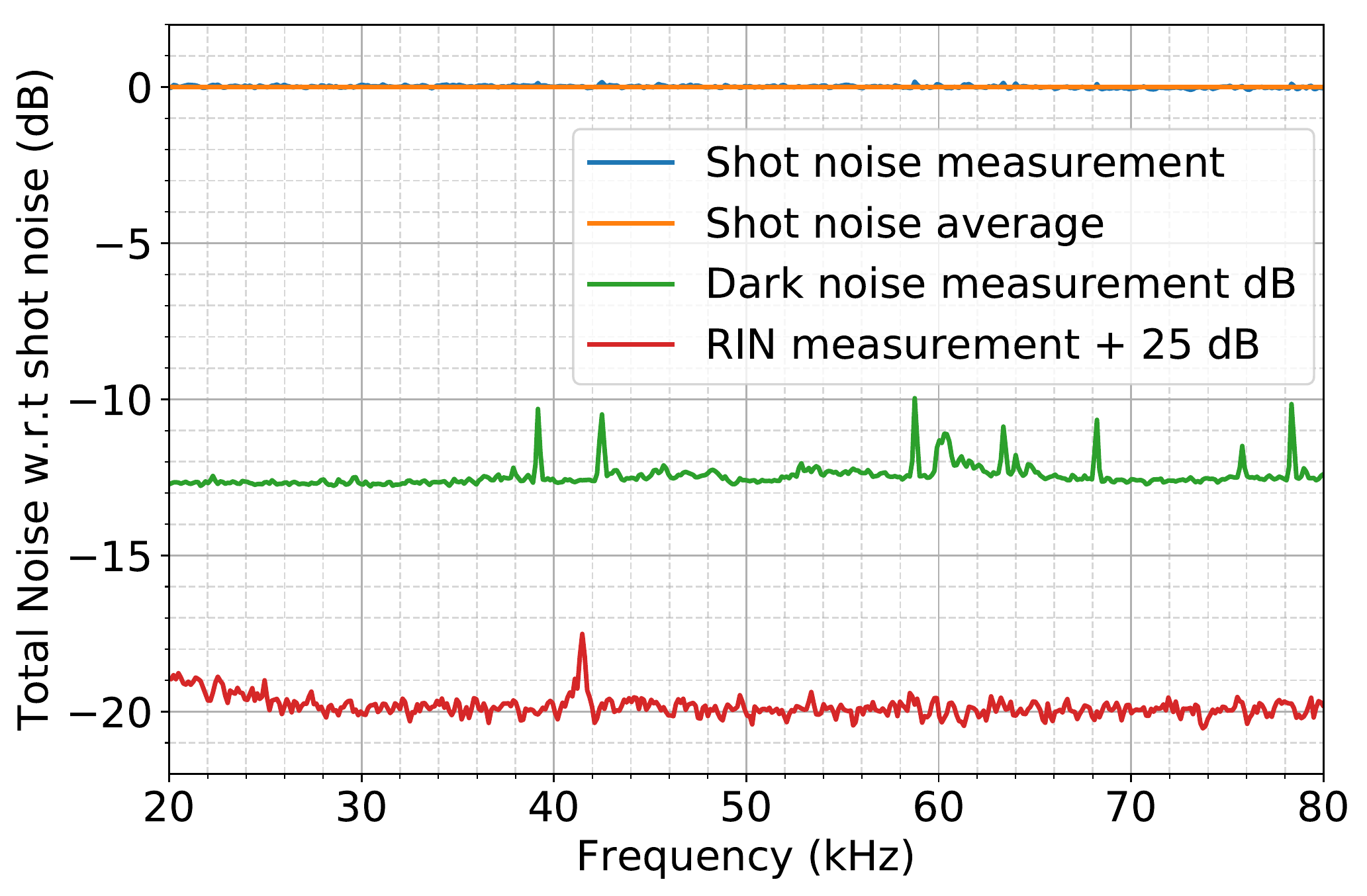}
	\caption{Classical laser intensity noise and dark noise, shown relative to shot noise. Since we always keep the total detected power on $\mathrm{PD_{sqz}}$ constant (and just change the LO power to change the measurement quadrature), the relative dark noise and classical laser intensity noise can just be scaled to that power. In this figure, 25 dB is added to the measured RIN so that it is viewable on the same graph as the other noise measurements.}
	\label{fig:ShotNoise}
\end{figure}

\subsection{Correlations}

Consider splitting an intensity squeezed beam onto two photodetectors. For convenience, let's split it as 50\%. The amplitude quadrature of the two fields hitting the photodetectors may then be written as
\begin{eqnarray}
a_1 = \frac{e^{-r}x_1 +c- y_1}{\sqrt{2}}\\
b_1 = \frac{e^{-r}x_1 +c+ y_1}{\sqrt{2}},
\end{eqnarray}
where $x_1$ is the vacuum that has been squeezed by the factor $e^{-r}$, $c$ represents any classical noise that might be present, and $y_1$ is the vacuum that enters at the beamsplitter.

If we measure the the averaged cross power spectrum of the resulting photocurrents, but don't take the absolute value, we find
\begin{equation}
\left<S_{ab}\right> = \frac{1}{2}\left(e^{-2r}+S_c-1\right)\alpha \beta,\label{eq:Sab}
\end{equation}
where we have normalized shot noise to 1, and assumed detector a has a relative gain of $\alpha$, and detector b has a relative gain $\beta$, and $S_c$ is the power spectrum of the classical noise scaled to shot noise. All the cross terms between $x_1$, $y_1$ and $c$ will average to 0, as they  are uncorrelated. If the original field is squeezed, then that requires $e^{-2r} + S_c < 1$, which would then imply $\left<S_{ab}\right> < 0$. Note that if this is not satisfied, such that we have classical noise that destroys the squeezing, then $e^{-2r} + S_c > 1$, which requires $\left<S_{ab}\right> > 0$. Therefore, by looking at the sign of the average cross power spectrum, one can definitively prove whether squeezing is present or not.

To interpret this, when the beam is limited by classical noise, the power fluctuations hitting both PDs are identical and positively correlated. If the beam is exactly shot noise limited, the power fluctuations hitting the two PDs are uncorrelated. With a perfectly amplitude squeezed beam, the power fluctuations are exactly anti-correlated.

We may write the individual power spectra as
\begin{eqnarray}
\frac{S_a}{\alpha^2} = \frac{S_b}{\beta^2} = \frac{e^{-2r} + S_c + 1}{2}.
\end{eqnarray}
Then define the normalized correlation as
\begin{equation}
C=\frac{\left<S_{ab}\right>}{\sqrt{S_a S_b}} = \frac{e^{-2r} + S_c - 1}{e^{-2r}+S_c+1}.\label{eq:C}
\end{equation}
This is convenient because it supplies a unitless measure of the nature of the noise, and is independent of the relative gain of the photodetectors. This $C$ is similar to the square root of coherence, but retains phase information. In fact, the coherence may be written as $C C^*$.

We can see that if the field is entirely classical so that $S_c$ dominates, then $C=+1$. Likewise if the beam is exactly shot noise limited without squeezing, then $C=0$. Finally, for an infinitely squeezed field with no classical noise, $C=-1$.

\begin{figure}
	\centering
	\includegraphics[width=0.5\linewidth]{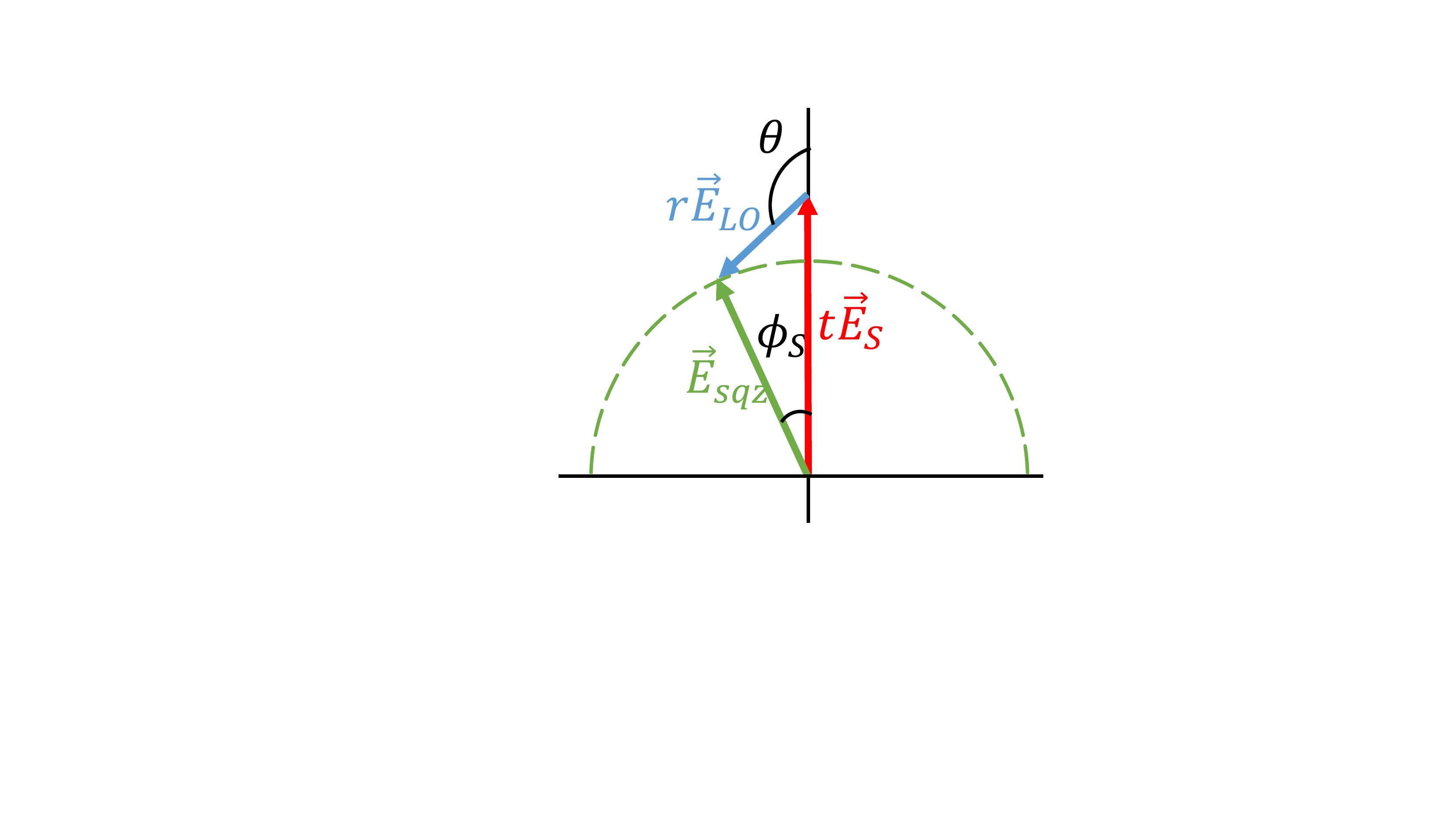}
	\caption{A phasor diagram showing how the tunable homodyne detector selects the measurement quadrature. The sum of the LO field (blue) and the signal field (red) selects the quadrature that is being measured (green). In the entire manuscript, we report this angle $\phi_{\rm S}$ as the measurement quadrature. We determine the quadrature by knowing the power in all the three fields, and the visibility. The dashed green circle represents a contour of constant detection power. In order to keep the shot noise reference unchanged, we choose to always lock $\mathrm{PD_{sqz}}$ with a constant total detected power, and vary the LO power to change the measurement quadrature. This has the effect of changing the angle $\theta$ of the LO.}
	\label{fig:phasor}
\end{figure}

To simplify,  let's call the total noise PSD of the original beam relative to shot noise $R = e^{-2r}+S_c$, in which case
\begin{eqnarray}
C=\frac{\left<S_{ab}\right>}{\sqrt{S_a S_b}} = \frac{R - 1}{R+1}.
\end{eqnarray}
This leads to
\begin{equation}
R = \frac{1+C}{1-C}. \label{eq:C_to_R}
\end{equation}
Thus, by measuring $C$, we have a method to measure the amount of noise relative to shot noise independent of our ability to calibrate shot noise. 

This treatment is simplified by not propagating the DC carriers of the fields. The final physical result becomes invariant of the beam splitter convention if one keeps track of the DC carrier fields. The cross spectrum $\left<S_{ab}\right>$ is negative for a squeezing beam $x$, irrespective of the beam splitter convention, as long as the carrier of the field $y$ is smaller than the carrier of the field $x$. This condition is trivially satisfied in our measurement, since $y$ is coming from vacuum fluctuations, with a zero DC field.

To include the effects of uncorrelated electronics noise on the photodetectors, we may rewrite the power spectrums for each detector as
\begin{eqnarray}
S_a = \alpha^2 \frac{R + 1}{2} + \alpha^2 S_{da}\\
S_b = \beta^2 \frac{R+ 1}{2} + \beta^2S_{db},
\end{eqnarray}
where $S_{da}$ and $S_{db}$ are the PSDs of each detector from electronics noise. The resulting normalized correlation is
\begin{align}
C&=\frac{\left<S_{ab}\right>}{\sqrt{S_a S_b}} = \frac{R - 1}{\sqrt{\left(R+1+S_{da}\right)\left(R+1+S_{db}\right)}}\nonumber\\
&= \frac{R-1}{R+1}\left[\sqrt{1+\frac{S_{da}}{1+R}}\sqrt{1+\frac{S_{db}}{1+R}}\right]^{-1/2}\nonumber\\
&=\eta \frac{R-1}{R+1},\label{eq:efficiency}
\end{align}
where $\eta\leq 1$ is an effective efficiency of the measurement. In our experiment, since the dark noise is far below shot noise (\cref{fig:ShotNoise}), the efficiency $\eta$ is close to 1. If instead one had a lower efficiency, we can see from \cref{eq:efficiency} that electronics noise will make observed correlations trend towards 0, and the inferred quantum noise level to shot noise. 

\begin{figure}
	\includegraphics[width=\linewidth]{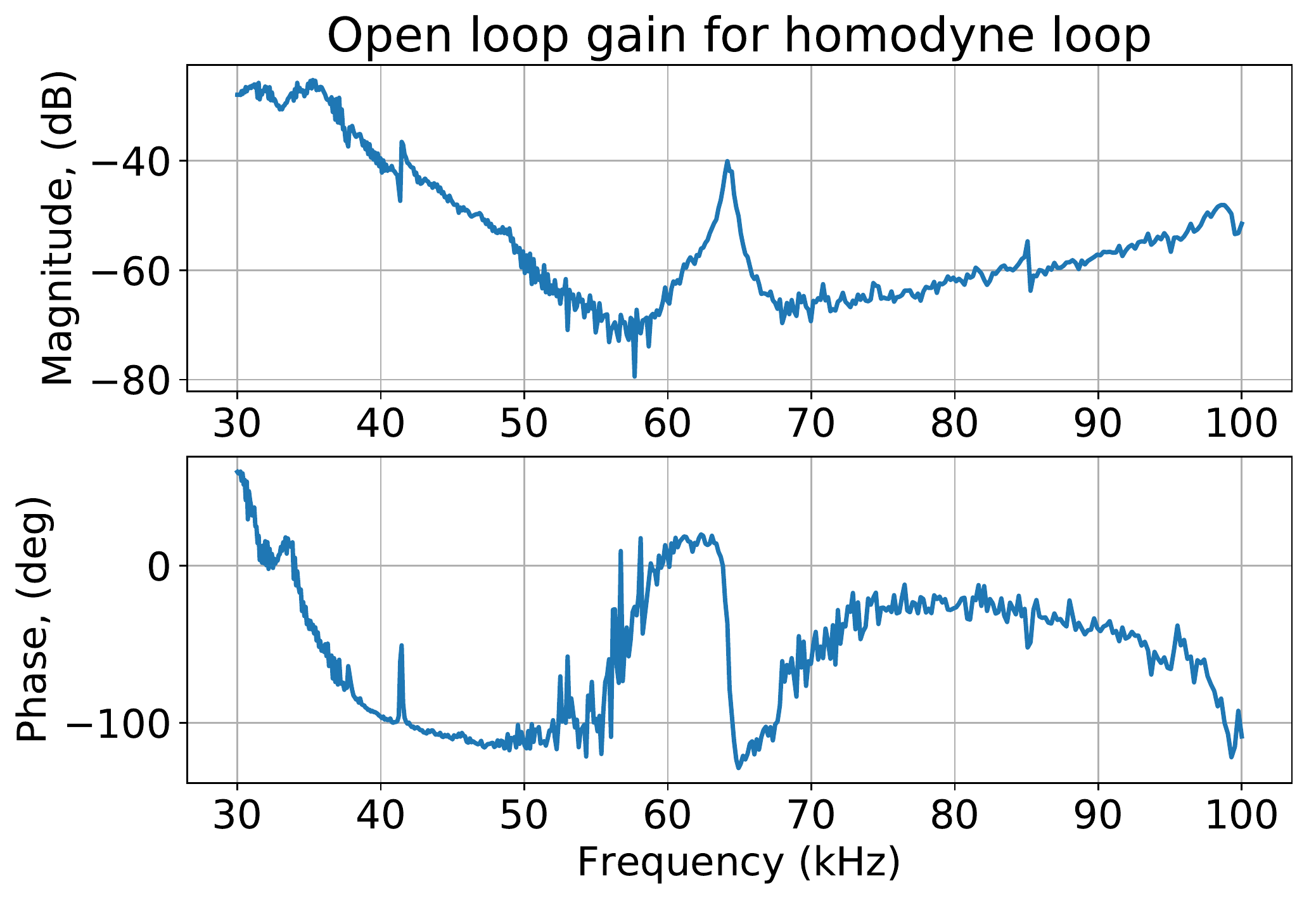}
	\caption{Measurement of the open loop transfer function of the homodyne locking loop around the squeezing frequency band. Since this loop is designed only to suppress large path length fluctuations between the local oscillator and the signal at low frequencies ($<$1 kHz), this loop has close to zero gain at our measurement frequencies.}
	\label{fig:HomodyneLoop}
\end{figure}

\subsection{Tunable homodyne detector}

In our experiment, we opt to use a single-photodiode homodyne detection. Instead of combining the signal beam with the local oscillator (LO) on a 50-50 beam splitter, we combine it on a highly asymmetric beam splitter. We measure on the output port which transmits the larger signal fraction and reflects the smaller fraction of the LO. While this scheme introduces some loss of signal, it works with just a single photodetector and eliminates the need for performing perfect subtraction that is needed in a balanced homodyne. Since the amount of squeezing expected in this experiment is relatively low, the reduction in squeezing due to this beam splitter loss is small.

We show below that the signal quadrature in which the measurement is performed is given by the angle made by the resultant of vector addition of the carrier of signal and LO with respect to  the signal, as displayed in \cref{fig:phasor}. Similarly, the LO quadrature measured is given by the angle this resultant makes with the LO.

\begin{subequations}
	\begin{gather}
		\vec{E}_\mathrm{sqz} = t \vec{E}_\mathrm{S} + r \mathbb{R}(\theta)\, \vec{E}_\mathrm{LO}\\
		\tan\phi_{\rm S} = \frac{r E_{\rm LO}\sin\theta}{rE_{\rm LO}\cos\theta+tE_{\rm S}}\\
		\tan\phi_{\rm LO} = \frac{-tE_{\rm S}\sin\theta}{rE_{\rm LO} + tE_{\rm S}\cos\theta}\\
		\vec{E}_\mathrm{sqz} = |\vec{E}_\mathrm{sqz}|\, \vec{U}(\phi_{\rm S})
	\end{gather}\label{eq:MeasurementQuadrature}
\end{subequations}

Here $\vec{E}$ represents the strength of the carrier of the signal(S), LO and the resultant (sqz). $t$ is amplitude transmitivity ($\sqrt{0.965}$) and $r$ is amplitude reflectivity ($\sqrt{0.035}$) of the beam splitter. We define a unit vector $\vec{U}(\phi_{\rm S})$ which represents a vector in the direction of the resultant, and determines the measured quadrature.
Using \cite{Corbitt_mathematical}, we can also calculate the loss effect of the beam splitter. We define $\vec{e}$ as the fluctuations on the field, normalized such that the shot noise is $|\vec{E}|^2$ \cite{Aggarwal_ClosedLoop}. We then propagate these fluctuations from the signal and the LO as they interfere on the beam splitter : 

\begin{subequations}
	\begin{align}
		\vec{e}_{\rm sqz} = ~&t\vec{e}_{\rm S} + r{\rm e}^{i \phi} \mathbb{R}(\theta)\, \vec{e}_{\rm LO}\\
		\vec{E}_{\rm sqz}.\vec{e}_\mathrm{sqz} = ~&|\vec{E}_\mathrm{sqz}| \,\vec{U}(\phi_{\rm S})^\dagger.\vec{e}_\mathrm{sqz}\\
		S_\mathrm{sqz}(\phi_{\rm S}) = ~&\,\vec{U}(\phi_{\rm S})^\dagger. \vec{e}_{\rm sqz}.\vec{e}^\dagger_{\rm sqz} .\vec{U}(\phi_{\rm S})\\
		= ~&\,\vec{U}(\phi_{\rm S})^\dagger.
		( t^2\vec{e}_{\rm S}\vec{e}^\dagger_{\rm S} \,+\\
		&r^2 \mathbb{R}(\theta)\, \vec{e}_{\rm LO}\vec{e}^\dagger_{\rm LO}\mathbb{R}(\theta)^\dagger )
		.\vec{U}(\phi_{\rm S})\nonumber\\
		= ~&t^2 S_\mathrm{S}(\phi_{\rm S}) + r^2
	\end{align}
\end{subequations}
where we have assumed that the LO is shot noise limited, and defined $S_\mathrm{S}(\phi_{\rm S}) = \vec{U}(\phi_{\rm S})^\dagger \cdot \vec{e}_{\rm S}\vec{e}^\dagger_{\rm S} \cdot \vec{U}(\phi_{\rm S})  $ as the spectral density of the signal if measured perfectly with a balanced homodyne detector at the quadrature $\phi_{\rm S}$. The above equations show that the total spectral density measured, $S_\mathrm{sqz}(\phi_{\rm S})$ is a combination of $S_\mathrm{S}(\phi_{\rm S})$ and 1, in the ratio of the beam splitter's reflectivity.

%

\end{document}